\documentclass[12pt]{iopart}
\usepackage{iopams}
\usepackage{bbm,graphicx,color}

\begin{document}

\title[Critical behavior of number-conserving cellular
automata]{Critical behavior of number-conserving cellular automata
  with nonlinear fundamental diagrams}

\author{Henryk Fuk\'s}

\address{Department of Mathematics\\
         Brock University\\
         St. Catharines, Ontario L2S 3A1, Canada\\
         email: {\texttt hfuks@brocku.ca}}
\begin{abstract}
  We investigate critical properties of a class of number-conserving
  cellular automata (CA) which can be interpreted as deterministic
  models of traffic flow with anticipatory driving.  These rules are
  among the only known CA rules for which the shape of the fundamental
  diagram has been rigorously derived. In addition, their fundamental
  diagrams contain nonlinear segments, as opposed to majority of
  number-conserving CA which exhibit piecewise-linear diagrams.  We found
  that the nature of singularities in the fundamental diagram of these
  rules is the same as for rules with piecewise-linear diagrams.  The
  current converges toward its equilibrium value like $t^{-1/2}$, and
  the critical exponent $\beta$ is equal to $1$. This supports the
  conjecture of universal behavior at singularities in
  number-conserving rules.  We discuss properties of phase transitions
  occurring at singularities as well as properties of the intermediate
  phase.
\end{abstract}



\maketitle

\section{Introduction}

Cellular automata (CA) can be viewed as cells in a regular lattice
updated synchronously according to a local interaction rule, where the
state of each cell is restricted to a finite set of allowed values.
An interesting subclass of CA consists of rules possessing an additive
invariant. The simplest of such invariants is the total number of
sites in a particular state. CA with such invariant, often called
``conservative CA'' or ``number-conserving CA'', can be viewed as a
system of interacting and moving particles, where in the case of a
binary rule, 1's represent sites occupied by particles, and 0's
represent empty sites.

In a finite system, the flux or current of particles in equilibrium
depends only on their density, which is invariant. The graph of the
current as a function of density characterizes many features of the
flow, and is therefore called the fundamental diagram. For a majority
of conservative CA rules, fundamental diagrams are piecewise-linear,
usually possessing one or more ``sharp corners'' or singularities.
There exist a strong evidence of universal behavior at singularities,
as reported in \cite{paper19}.

Conservative CA appear in various applications, and some special cases
have been studied extensively. Rule 184, which is a discrete version
of the totally asymmetric exclusion process, is an example of such
special case
\cite{Krug88,Nagatani95,Nagel96,Belitsky98,paper11,
  NishinariT98,BelitskyKNS01,Blank03}.  Although much is known about
this particular rule, dynamics of other conservative rules exhibits
many features which are not fully understood, and more general results
are just starting to appear
\cite{Pivato02,Moreira03,Durand2003,Formenti2003,Morita2001}.  In
particular, there exists no general result explaining the shape of
fundamental diagrams for conservative Ca, although promising results
have been obtained for some special cases by Kujala and Lukka
\cite{KujalaL02}, who studied a class of deterministic traffic 
rules introduced by \cite{paper5}. Using their results, we will study numerically
behavior of singularities for number-conserving rules with fundamental
diagrams different that previously reported.

It should be emphasized at this point that all CA rules considered
in this paper are strictly deterministic. Nevertheless, they
are strongly related to stochastic CA models of road traffic
flow, which have gained  widespread attention in recent years \cite{chow2000,helbing01}.

\section{Number-conserving cellular automata}

In what follows, we will assume that the dynamics takes place on
one-dimensional lattice of length $L$ with periodic boundary
conditions. Let $s_i(t)$ denote the state of the lattice site $i$ at
time $t$, where $i\in \{ 0, 1, \ldots, L - 1 \}$, $t \in \mathbbm{N}$.
All operations on spatial indices $i$ are assumed to be modulo $L$. We
will further assume that $s_i(t) \in \{ 0, 1\}$, and we will say that
the site $i$ is occupied (empty) at time $t$ if $s_i ( t ) = 1$ ($s_i
( t ) = 0$).

Let $l$ and $r$ be two integers such that $l \leq 0 \leq r$, and let
$n =r-l+1$. The set $\{ s_{i + l} ( t ), s_{i + l + 1} ( t ), \ldots,
s_{i + r} ( t )\}$ will be called the \textit{neighbourhood} of the
site $s_i ( t )$. Let $f$ be a function $f : \{ 0, 1 \}^n \rightarrow
\{ 0, 1 \}$, also called a \emph{local function} The update rule for
the cellular automaton is given by
\begin{equation}
  \label{cadef} s_i ( t + 1 )_{} = f ( s_{i + l} ( t ), s_{i + l + 1} ( t ),
  \ldots, s_{i + r} ( t ) ) .
\end{equation}
The CA rule given by (\ref{cadef}) will be called \emph{number conserving}
if
\begin{equation}
  \label{conscond} \sum_{i = 0}^{L - 1} f ( x_{i + l}, x_{i + l + 1}, \ldots,
  x_{i + r} ) = \sum_{i = 0}^{L - 1} x_i
\end{equation}
for all $\{ x_{0,} x_1, \ldots, x_{L - 1} \} \in \{ 0, 1 \}^L$. Note
that the above condition simply states that the number of occupied
sites is constant in time.

In \cite{Hattori91}, Hattori and Takesue demonstrated that CA rules
are number-conserving if and only if a discrete version of a standard
current conservation law $\partial \rho / \partial t = - \partial j /
\partial x$ is satisfied. More precisely, CA rule $f$ is
number-conserving if and only if for all $\{ x_1, x_2, \ldots, x_n \}
\in \{ 0, 1 \}^n$ it satisfies
\begin{equation}
  \label{curcons} f ( x_1, x_2, \ldots, x_n ) - x_{- l + 1} = J ( x_1, x_2,
  \ldots, x_{n - 1} ) - J ( x_2, x_3, \ldots, x_n ),
\end{equation}
where
\begin{equation}\label{curdef}
    J(x_1,x_2,\ldots,x_{n-1})=
    - \sum_{k=1}^{n-1}
f(\underbrace{0,0,\ldots,0}_k,x_1,x_2,\ldots,x_{n-k}) +
\sum_{j=1}^{-l} x_j.
\end{equation}

Applying this to all lattice sites we obtain
\begin{equation} \label{allsitescons}
    f(s_{i+l},\ldots,s_{i+r})-
    s_{i}=J(s_{i+l},\ldots,s_{i+r-1})
    -J(s_{i+l+1},\ldots,s_{i+r}),
\end{equation}
where we dropped $t$ dependence for clarity.

The above equation can be interpreted in a similar way as a
conservation law in a continuous, one dimensional physical system. In
such system, let $\rho(x,t)$ denote the density of some material at
point $x$ and time $t$, and let $j(x,t)$ be the current (flux) of this
material at point $x$ and time $t$. A conservation law states that the
rate of change of the total amount of material contained in a fixed
domain is equal to the flux of that material across the surface of the
domain. The differential form of this condition can be written as
\begin{equation}\label{concons}
    \frac{\partial \rho}{\partial t}=-\frac{\partial j}{\partial
    x}.
\end{equation}

Interpreting $s_i(t)$ as the density, the left hand side of
(\ref{allsitescons}) is simply the change of density in a
single time step, so that (\ref{allsitescons}) is an obvious
discrete analog of the current conservation law (\ref{concons})
with $J$ playing the role of the current.

Let us now assume that the initial distribution $\mu$ is a Bernoulli
distribution, i.e., at $t=0$, all sites $s_i(t)$ are independently
occupied with probability $p$ or empty with probability $1-p$, where
$p\in [0,1]$. Let us define
\begin{equation}
\rho(i,t)=E_\mu (s_i(t)).
\end{equation}
Since the initial distribution is $i$-independent, we expect that
$\rho(i,t)$ also does not depend on $i$, and we will therefore define
$\rho(t)=\rho(i,t)$.  Furthermore, for conservative CA, $\rho(t)$ is
$t$-independent, so we define $\rho=\rho(t)$. For the aforementioned
Bernoulli distribution we thus obtain $\rho=p$.  We will refer to
$\rho$ as the density of occupied sites. The expected value of the
current $J(s_{i+l}(t),s_{i+l+1}(t),\ldots,s_{i+r-1}(t))$ will also be
$i$-independent, so we can define the expected current as
\begin{equation} \label{defexpcurrent}
j(\rho,t)=E_\mu
\big(J(s_{i+l}(t),s_{i+l+1}(t),\ldots,s_{i+r-1}(t))\big).
\end{equation}

\begin{figure}
\begin{center}
\includegraphics[scale=0.55]{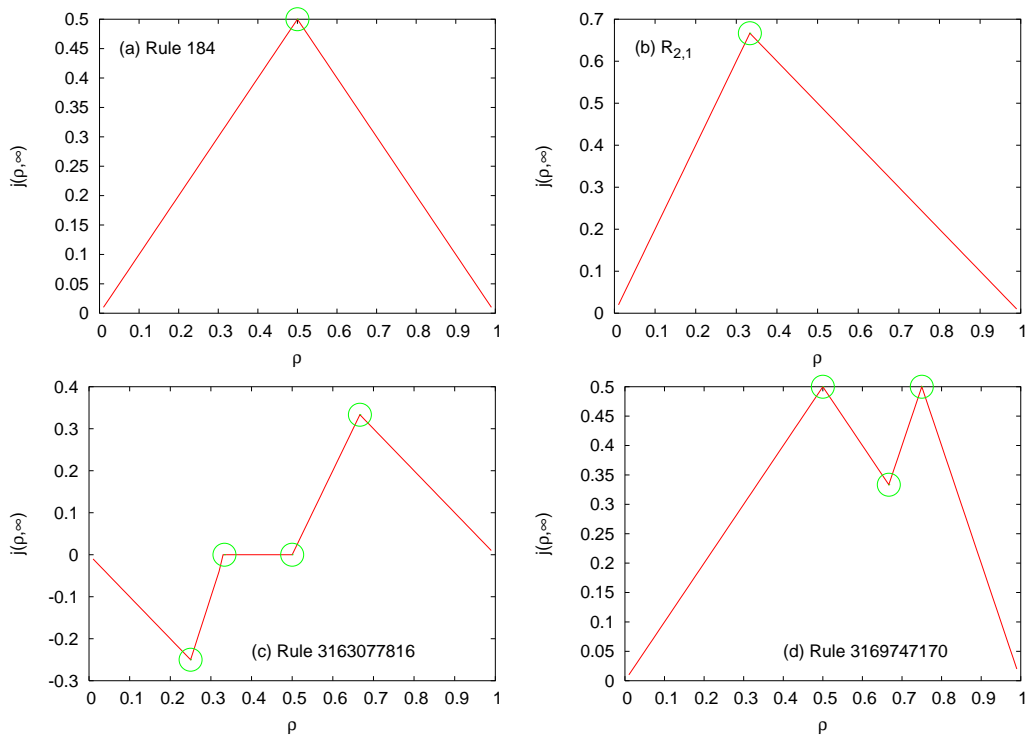}
\end{center}
\caption{Examples of fundamental diagrams for number-conserving
  rules. Singularities are denoted by green circles.  Diagrams (c) and
  (d) represent $n=5$ rules with indicated Wolfram's numbers.}
\label{examplefd}
\end{figure}

The graph of the equilibrium current $j(\rho,\infty)=\lim_{t
  \rightarrow \infty} j(\rho,t)$ versus the density $\rho$ is known as
the fundamental diagram. It has been numerically demonstrated
\cite{paper8} that for conservative deterministic CA the fundamental
diagram usually develops a singularity as $t\rightarrow \infty$,
meaning that $j(\rho,\infty)$ is not everywhere differentiable
function of $\rho$. A well-know example is CA rule 184, for which
\begin{equation}
s_i(t+1)=f\Big(s_{i-1}(t),s_i(t),s_{i+1}(t)\Big),
\end{equation}
and $f$ is defined by
\begin{eqnarray} \label{r184-01}
& & f(0,0,0)=0,\,  f(0,0,1) = 0,\,  f(0,1,0) = 0, \nonumber\\
& & f(0,1,1)=1,\, f(1,0,0)= 1,\, f(1,0,1) = 1, \nonumber\\
& & f(1,1,0)= 0,\, f(1,1,1) = 1.
\end{eqnarray}
The above definition can also be written in a form  (\ref{allsitescons}) as
\begin{eqnarray} \label{r184-02}
s_{i}(t+1)=s_i(t) + J(s_{i-1}(t), s_i(t)) - J(s_{i}(t), s_{i+1}(t)),
\end{eqnarray}
where $J(x_1,x_2)=x_1 (1-x_2)$.  The graph of the equilibrium current
for this rule is shown in Figure~\ref{examplefd}a. The singularity
appears at $\rho=0.5$.
Quite often, fundamental diagrams of number-conserving rules consist
of a finite number of linear segments separated by similar
singularities, as shown in Figure~\ref{examplefd}b--d.

Since number-conserving CA rules conserve the number of occupied
sites, we can label each occupied site (or ``particle'') with an
integer $n\in\mathbbm{Z}$, such that the closest particle to the right
of particle $n$ is labeled $n+1$. If $y_n(t)$ denotes the position of
particle $n$ at time $t$, the configuration of the particle system at
time $t$ is described by the increasing bisequence
$\{y_n(t)\}_{n=-\infty}^{\infty}$.  We can then specify how the
position of the particle at the time step $t+1$ depends on positions
of the particle and its neighbours at the time step $t$. For example,
for rule 184 one obtains
\begin{equation} \label{r184-00}
y_n(t+1)= y_n(t) + \min\{y_{n+1}(t) - y_n(t)-1,1\}.
\end{equation}

Equation (\ref{r184-00}) is sometimes referred to as the motion
representation. For arbitrary number-conserving CA rule it is possible
to obtain the motion representation by employing algorithm described
in \cite{paper10}.  The motion representation is analogous to Lagrange
representation of the fluid flow, in which we observe individual
particles and follow their trajectories.  On the other hand, eq.
(\ref{r184-02}) could be called Lagrange representation, because it
describes the process at a fixed point in space. The Euler-Lagrange
analogy has been explored in \cite{MatsukidairaN03}.

In practice, the choice between Euler and Langrange description of the
number-conserving CA is usually dictated by convenience. Cellular
automata rules which we want to consider in this paper are easier to
define using the Euler paradigm, as we will see in the next section.

\section{Anticipatory driving rules}
In what follows, we will consider a class of CA rules first introduced
in \cite{paper5}. They are defined in terms of two positive integers
$m$ (maximum speed) and $k$ (blocking parameter).  These rules can be
interpreted as simplified deterministic models of road traffic flow with driver's
anticipation. 
Occupied sites represent cars on a single-lane road.
All cars move synchronously using the following algorithm. Driver of
each car first locates the nearest gap (cluster of empty sites) in
front of him, the length of this gap denoted by $g$. If the first
empty site (i.e., the first site belonging to the aforementioned gap)
is further than $k$ sites ahead, the car does not move, otherwise, it
moves to the site $i+v$, where $v=\min (g,m)$.  More formally, using
Euler's paradigm, and denoting position of the $n$-th car by $y_n(t)$,
we can write
\begin{equation}\label{rdef1}
y_n(t+1)=y_n(t)+[[y_b(t)-y_n(t)+1 \leq k]]\min \{g,m\},
\end{equation}
where
\begin{eqnarray} \label{rdef2}
b&=&\min\{i\in \mathbbm{Z} :  i\geq n \mbox{\,\,and\,\,} y_{i+1}(t)-y_i(t)>1\},\\
\label{rdef3}
g&=&y_{b+1}(t)-y_b(t)-1,
\end{eqnarray}
and where $[[ P ]]=1$ if the statement $P$
is true, otherwise $[[ P ]]=0$.

Cellular automaton rule defined by (\ref{rdef1}--\ref{rdef3}) will be
referred to as as ${\cal R}_{m,k}$.  To illustrate dynamics of this
rule, let us assume that $m=k=2$. Consider, for example, the following
configuration: \vspace*{10pt}
\begin{center}
\begin{tabular}{|c|c|c|c|c|c|c|c|c|c|c|c|c|} \hline
$\cdots$&0&0&A&B&0&0&C&0&D&0&0&$\cdots$\\ \hline
$\cdots$&0&0&0&0&A&B&0&C&0&0&D&$\cdots$\\ \hline
\end{tabular}
\end{center}
\vspace*{10pt} The first row represents configuration of cars A, B, C,
D at time $t$, and the second line their configuration at time $t+1$.
Zeros denote empty sites. Car B sees ahead a gap of length 2, so it
will move by two sites.  Car A sees the same gap, so it will move by
two sites as well. As a result, cars A and B move together as if they
formed a single ``block''. This is called ``anticipatory driving'' --
car A anticipates that car B will move.  For rule ${\cal R}_{m,k}$,
cars can move in blocks of length up to $k$.

Compare this with $m=2$, $k=1$ case:
\vspace*{10pt}
\begin{center}
\begin{tabular}{|c|c|c|c|c|c|c|c|c|c|c|c|c|} \hline
$\cdots$&0&0&A&B&0&0&C&0&D&0&0&$\cdots$\\ \hline
$\cdots$&0&0&A&0&0&B&0&C&0&0&D&$\cdots$\\ \hline
\end{tabular}
\end{center}
Car A has not moved, because the first gap ahead of it begins further
than $k=1$ sites away, so A cannot see it, and, as a result, cannot
anticipate that B will move.

\section{Known results for ${\cal R}_{m,1}$ and
  ${\cal R}_{1,k}$ }
 For rules ${\cal R}_{m,1}$, $m=1,2,\ldots$, the
current $j(t,\rho)$, as defined by eq. (\ref{defexpcurrent}), can be
computed explicitly. As shown in \cite{paper11}, it is given by
\begin{equation} \label{FIflux}
j(\rho,t) =1-\rho- \sum_{i=1}^{t+1} \frac{i}{t+1} {{(m+1)(t+1)}
\choose
 {t+1-i}} \rho^{t+1-i} (1-\rho)^{m(t+1)+i}.
\end{equation}
One can then show that in the limit of $t\to\infty$, the equilibrium
current $j(\rho,\infty)$ is a piecewise linear function of $\rho$
given by
\begin{equation} \label{FIfluxInfty}
j(\rho,\infty)= \left\{ \begin{array}{ll}
 m \rho  & \mbox{if $\rho<1/(m+1)$}, \\
 1-\rho    & \mbox{otherwise},
\end{array}
\right.
\end{equation}
as shown in Figure~\ref{examplefd}(b).

For $\rho<1/(m+1)$ the average velocity of particles at equilibrium is
$m$, \textit{i.e.}, all particles are moving to the right with the
maximum speed $m$. The system is said to be in the \emph{a free-moving
  phase}. When $\rho>1/(m+1)$, the speed of some particles is less
than the maximum speed $m$. The system is in the so-called
\emph{jammed phase}.

The transition from the free-moving phase to the jammed phase occurs
at $\rho=\rho_c=1/(m+1)$ called the \emph{critical density}. At
$\rho_c$, it is possible to obtain an asymptotic approximation of
(\ref{FIflux}) by replacing the sum by an integral and using de
Moivre-Laplace limit theorem, as done in \cite{paper11}. At $\rho_c$
this procedure yields
\begin{equation}
j(\rho,\infty)-j(\rho,t) =\sqrt{\frac{m}{2 \pi (m+1)t}}
\left(e^{-\frac{(m+1)}{2 m t}} - e^{-\frac{(m+1) t}{2 m}} \right),
\end{equation}
and therefore
\begin{equation}
j(\rho,\infty)-j(\rho,t) \sim t^{-1/2}.
\end{equation}
Here,
by $f(t)\sim g(t)$ we mean that $\lim_{t\to\infty} f(t)/g(t)$
exists and is different from $0$.

Very similar results can be obtained for ${\cal R}_{1,k}$,
$k=1,2,\ldots$, if one takes advantage of the duality property.
Duality in this context means that if cars are moving to the right
according to the rule ${\cal R}_{m,k}$, then empty sites can be
treated as another type of particles which are moving to the left
according to ${\cal R}_{k,m}$. Therefore, if we replace $\rho$ by
$1-\rho$ in eq. (\ref{FIflux}), we will obtain the equilibrium current
for ${\cal R}_{1,k}$
\begin{equation} \label{FIfluxdual}
j(\rho,t) =\rho- \sum_{i=1}^{t+1} \frac{i}{t+1} {{(k+1)(t+1)}
\choose
 {t+1-i}} (1-\rho)^{t+1-i} \rho^{k(t+1)+i},
\end{equation}
and
\begin{equation} \label{FIfluxInftydual}
j(\rho,\infty)= \left\{ \begin{array}{ll}
 \rho  & \mbox{if $\rho<k/(k+1)$}, \\
 k(1-\rho)    & \mbox{otherwise}.
\end{array}
\right.
\end{equation}
As in the case of ${\cal R}_{m,1}$, rule ${\cal R}_{1,k}$ exhibits two
phases represented by linear segments of the fundamental diagram,
separated by the singularity at $\rho=k/(k+1)$.
\section{Rule ${\cal R}_{2,2}$}
For arbitrary $m$ and $k$ both greater than $1$, explicit expressions
for the equilibrium current are not known. Nevertheless, Kujala and
Lukka \cite{KujalaL02} presented an efficient algorithm which can
determine the steady-state current starting from a given initial
state.  Using that algorithm and the method of generating functions,
they obtained polynomial equations relating the equilibrium current
and the density.

When $m>1$ or $k>1$, the fundamental diagram for ${\cal R}_{m,k}$ is
different than in the case of ${\cal R}_{m,1}$ or ${\cal R}_{1,k}$.
In addition to the free moving phase and the jammed phase, a novel
phase appears, which we will call an \emph{intermediate phase}.  The
smallest values of $m$ and $k$ for which one can observe this
phenomenon is $m=k=2$, and we will use these values in subsequent
considerations.

For $m=k=2$, the method of \cite{KujalaL02} method yields
\begin{equation} \label{r22flux}
j(\rho,\infty)= \left\{ \begin{array}{ll}
 2\rho  & \mbox{if $\rho\leq \rho_{c1}$}, \\
 C  & \mbox{if $\rho_{c_1}<\rho<\rho_{c2}$}, \\
 2(1-\rho)  & \mbox{if $\rho \geq \rho_{c2}$},
\end{array}
\right.
\end{equation}
where
\begin{eqnarray}
\rho_{c1}&=&\frac{6}{7} -\frac{2\sqrt{2}}{7},\\
\rho_{c2}&=&1-\rho_{c1},
\end{eqnarray}
and where $C$ is a solution of
\begin{equation}
16 C^2+\rho^2 (1-\rho^2)(8 C^2-36  C^3) + (1+27 \rho^2 (1-\rho^2)) C^4 - C^5=0.
\end{equation}
\begin{figure}
\begin{center}
\includegraphics[scale=0.9]{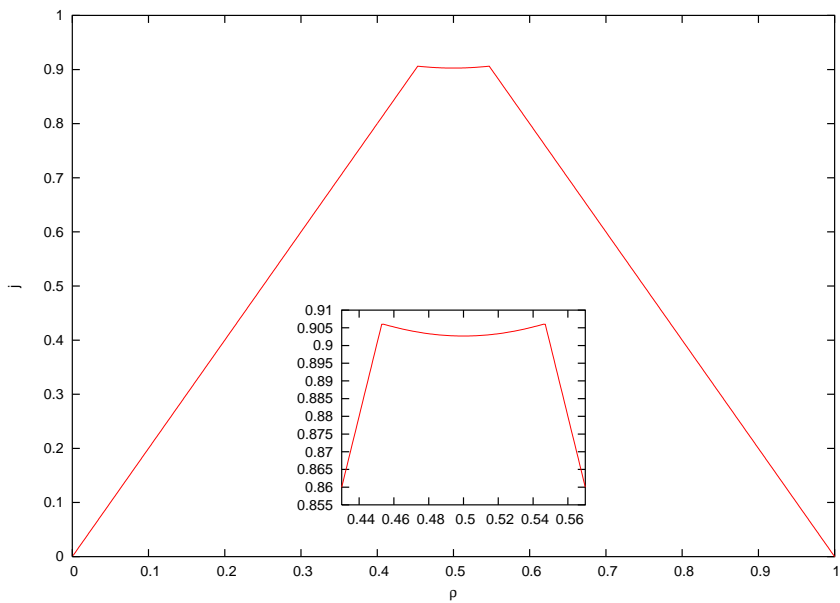}
\end{center}
\caption{Fundamental diagram for rule ${\cal R}_{2,2}$ as
given by eq. (\ref{r22flux}).
The inset shows magnification of the non-concavity region.}
\label{r22cur}
\end{figure}
Figure \ref{r22cur} shows the graph of $j(\rho, \infty)$ as a function
of $\rho$ for rule ${\cal R}_{2,2}$. Two singularities at $\rho_{c1}$
and $\rho_{c2}$ are clearly visible. By singularities we understand
values of $\rho$ for which $j(\rho, \infty)$ is non-differentiable
with respect to $\rho$. The two singularities divide the fundamental
diagram into three parts. When $\rho\leq \rho_{c1}$, all particles are
moving with velocity $2$, so we will call this a \emph{free moving
  phase}.  The region $\rho_{c_1}<\rho<\rho_{c2}$ will be called an
\emph{intermediate phase}, and the region $\rho \geq \rho_{c2}$, in
analogy to the fundamental diagram described by eq.
(\ref{FIfluxInfty}), will be called a \emph{jammed phase}.

The existence of singularities in fundamental diagrams of
number-conserving rules is not new, and their properties have been
documented in \cite{paper11,paper19}.  The most common singularity
type is a singularity which separates two linear segments of the
fundamental diagram, just like $\rho=1/(m+1)$ for ${\cal R}_{m,1}$
(eq.\ref{FIfluxInfty}). At all such singularities, numerical evidence
suggests that the relaxation to the equilibrium follows the same power
law, i.e.  $j(\rho,\infty)-j(\rho,t) \sim t^{-1/2}$.

The two singularities in rule ${\cal R}_{2,2}$ are different from the
singularity observed in ${\cal R}_{m,1}$, since they separate linear
segments of the fundamental diagram from the nonlinear segment. Yet
surprisingly, they exhibit the same power law behaviour, as we shall
see in the next section.

\section{Convergence to equilibrium}

\begin{figure}
\begin{center}
\includegraphics[scale=0.9]{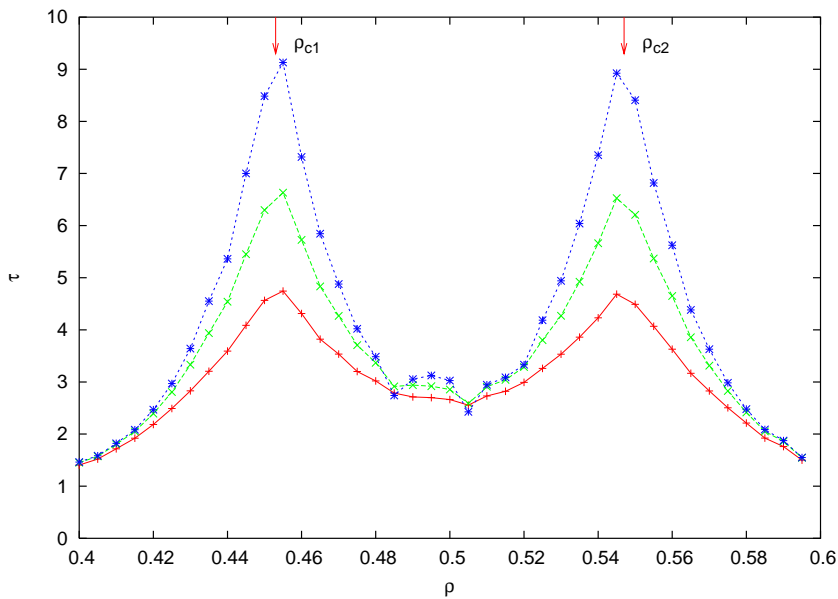}
\end{center}
\caption{Truncated decay time $\tau_T(\rho)$ as a function of $\rho$
  for $T=100$ ($+$), $T=200$ ($\times$) and $T=400$ ($\star$) for a
  lattice of $1000$ sites. Each point represents average of $100$
  runs.}
\label{taufigure}
\end{figure}

In order to define current for ${\cal R}_{2,2}$, we will first write
definition of ${\cal R}_{2,2}$ using Lagrange's representation. Its
local function $f$ appearing in eq. (\ref{cadef}) can be written as in
a compact form as
\begin{eqnarray}
f(x_1,x_2,x_3,x_4,x_5)=x_1 +x_2 x_4 -x_1 x_3
-x_1 x_2 x_4 -x_2 x_3 x_4 +x_2 x_3 x_5 \nonumber \\
 +x_1 x_2 x_3 x_4 -x_2 x_3 x_4 x_5  +x_3 x_4 x_5,
\end{eqnarray}
with $l=-2$, $r=2$. The local current, as defined by eq. (\ref{curdef}),
becomes
\begin{equation}
J(x_1,x_2,x_3,x_4)=x_1+x_2-x_1 x_2 x_4-x_1 x_3+x_1 x_2 x_3 x_4-x_2 x_3 x_4.
\end{equation}
In order to estimate the expected current $j(\rho,t)$ as defined by
eq. (\ref{defexpcurrent}), we performed a series of numerical
experiments.  We start with a lattice of $L=10^6$ sites, where
initially each site is occupied with probability $\rho$ and empty with
probability $1-\rho$. After $t$ iterations, the average current is
then given by
\begin{equation}
\langle j(\rho,t)\rangle=\frac{1}{L} \sum_{i=0}^{L-1} J(s_i(t),s_{i+1}(t),s_{i+2}(t),s_{i+3}(t)).
\end{equation}
\begin{figure}
\begin{center}
\includegraphics[scale=0.9]{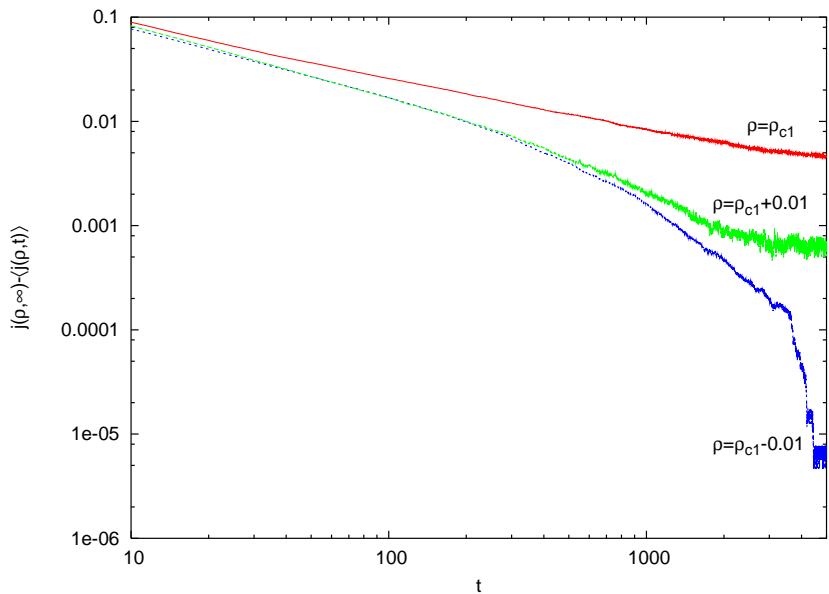}
\end{center}
\caption{Graphs of $ j(\rho,\infty)-\langle j(\rho,t)\rangle $ as a function of time for
  rule ${\cal R}_{2,2}$.}
\label{expcur}
\end{figure}
Figure~\ref{expcur} shows graphs of $j(\rho,\infty)-\langle
j(\rho,t)\rangle $ for three different values of the density $\rho$:
exactly at the singularity ($\rho=\rho_{c1}$), slightly below
($\rho=\rho_{c1}-0.01$), and slightly above ($\rho=\rho_{c1}+0.01$).
Due to the symmetry of the fundamental diagram of ${\cal R}_{2,2}$,
the behaviour in the vicinity of $\rho_{c_2}$ is the same, therefore
we are only considering $\rho_{c_1}$.  We can see that at the
singularity the power law behavior is observed, as evidenced by the
straight line in the log-log plot. This is not the case, however, for
densities smaller or grater than than $\rho_{c1}$.  For
$\rho=\rho_{c1}$, fitting the curve
\begin{equation}
j(\rho,\infty)-\langle j(\rho,t)\rangle = A t^\alpha
\end{equation}
to the data set visualized in Figure~\ref{expcur}, we obtain
$\alpha=-0.489 \pm 0.005$, which is very close to values reported
previously for singularities in other number-conserving rules, and
very close to the exact value $\alpha=-1/2$ known for ${\cal
  R}_{1,m}$. This suggests that singularities in ${\cal R}_{2,2}$ have
the same nature as singularities in other number-conserving rules. To
confirm this observation, we will introduce the notion of the
\emph{decay time} defined as
\begin{equation} \label{tau}
\tau(\rho) = \sum_{t=0}^{\infty} j(\rho,\infty)-j(\rho,t).
\end{equation}
If the decay is of power-law type, i.e.  $j(\rho,\infty)-\langle
j(\rho,t)\rangle \sim t^\alpha$ with $\alpha \leq 1$, then the
infinite sum in eq. (\ref{tau}) should diverge. If, on the other hand,
the decay is exponentially fast, we should see rapid convergence. For
five-input ($n=5$) number-conserving CA rules with piecewise-linear
fundamental diagrams investigated in \cite{paper19}, the divergence of
$\tau(\rho)$ has been observed only at the singularity, while away
from the singularity $\tau(\rho)$ it quickly converged.
Figure~\ref{taufigure} shows graphs of the truncated decay time
defined as
\begin{equation} \label{tausim}
\tau_T(\rho) = \sum_{t=0}^{T} j(\rho,\infty)-j(\rho,t)
\end{equation}
as a function of $\rho$ for several values of $T$.  One can see that
the decay time diverges at critical points $\rho_{c1}$ and
$\rho_{c2}$, similarly as reported in \cite{paper19} for other
number-conserving rules.

\section{Phase transitions}
Divergence of the decay time at the critical point $\rho_{c1}$ is
somewhat similar to the critical slowing down observed in phase
transitions -- the closer to the singularity, the longer it takes to
reach the steady state. In fact, it is possible to define the order
parameter for rule ${\cal R}_{2,2}$, thus interpreting the singularity
at $\rho_{c1}$ as the critical point in a second-order kinetic phase
transition. 

We will first note that the maximum possible current occurs when all
particles are moving with the maximum speed, which for ${\cal
  R}_{2,2}$ is $2$, hence the maximum current is equal to $2 \rho$.
The natural choice of the order parameter
would be therefore the
difference between the maximum possible current and the actual current
in the steady state\footnote{Note that this is not an order parameter in the strict sense,
i.e., there is no obvious symmetry-breaking in the ordered phase.}. 
We will denote this order parameter by $M(\rho)$,
formally defined as
\begin{equation}
M(\rho)=2 \rho - j(\rho,\infty).
\end{equation}
The order parameter is zero in the free-moving phase, and becomes
non-zero in the intermediate phase. 

 Using eq. (\ref{r22flux}), it is
possible to obtain series expansion of $M(\rho)$ around $\rho_{c1}$
for $\rho \geq \rho_{c1}$:
\begin{equation}
M(\rho) \propto (10 \sqrt{2} -12) (\rho-\rho_{c1})
+ \frac{1}{2} (47 \sqrt{2} -69)
(\rho-\rho_{c1})^2 + O((\rho-\rho_{c1})^3).
\end{equation}
The critical exponent beta, normally defined by $M(\rho) \propto
(\rho-\rho_{c1})^\beta$, is therefore equal to $1$. Again, this is in
agreement with the value of $\beta$ obtained for other
number-conserving rules \cite{paper19}.

\section{Intermediate phase}
In spite of all similarities to other number-conserving rules,
the behavior of ${\cal R}_{2,2}$ is somewhat unusual, especially
in the intermediate phase, where the fundamental diagram is nonlinear.

As mentioned earlier, dynamics of number-conserving CA rules somewhat
resembles dynamics of the kinematic wave equation, which describes
propagation of density waves in a continuous medium.  For the
kinematic wave equation, the slope of the fundamental diagram at a
given point represents velocity of density waves. It turns out that
the slope of the fundamental diagram for number-conserving CA can be
interpreted in a similar way.

\begin{figure}
\begin{center}
(a) \includegraphics[scale=0.9]{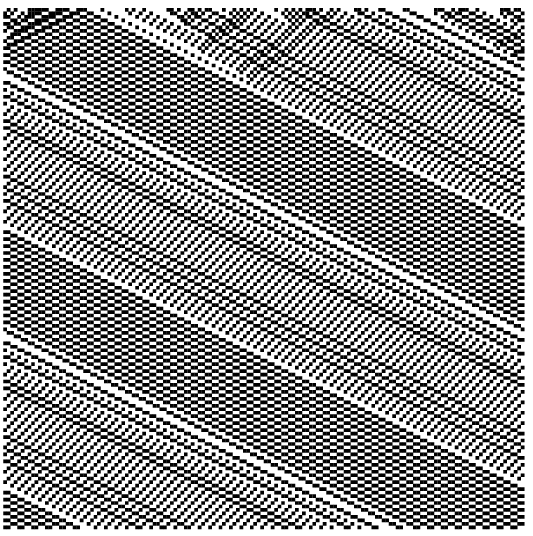}
\includegraphics[scale=0.9]{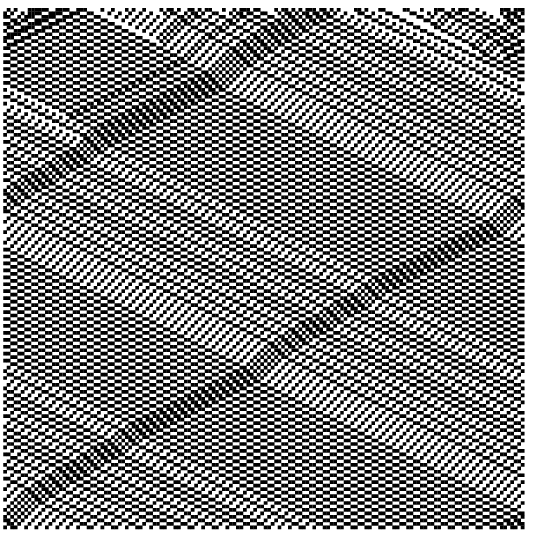} (b)\\
(c)\includegraphics[scale=0.9]{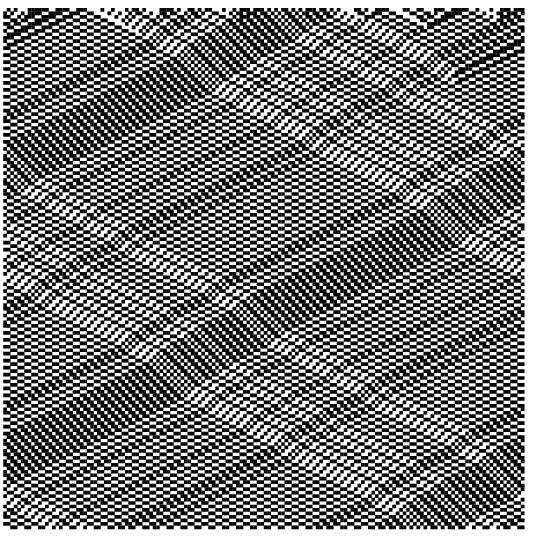}
\includegraphics[scale=0.9]{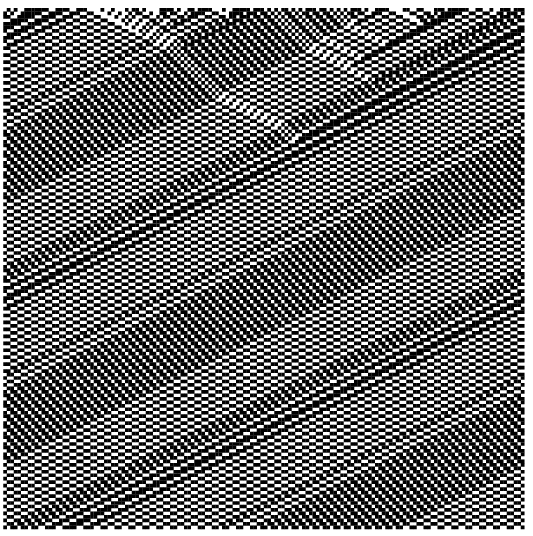} (d)
\end{center}
\caption{Spatio-temporal patterns generated by ${\cal R}_{2,2}$
  for random initial configurations with densities (a) $\rho=0.42$,
  (b) $\rho=0.47$, (c) $\rho=0.53$, (d) $\rho=0.58$.  Vertical axis
  (increasing downward) represents time, while horizontal axis
  represents space. Blacks squares represent occupied sites.}
\label{patterns}
\end{figure}

In the free moving phase, the slope of the fundamental diagram given
by eq. (\ref{r22flux}) is equal to $\frac{\partial j}{\partial
  \rho}=2$. Compare this to Figure~\ref{patterns}(a) showing the
spatio-temporal diagram for $\rho=0.42$, which is in the free-moving
phase. If we treat regions of alternating blocks $11$ and $00$ as the
``background'' $\ldots 0011001100110011\ldots$, then white structures
can be clearly identified in that background. They propagate to the
right with velocity $2$. These are analogs of density waves -- in
fact, they are regions of local density smaller than $1/2$, i.e.,
blocks of zeros longer than 2 or block of ones shorter than 2. Free
moving phase is dominated by such density waves.  On the other hand,
in the jammed phase, only density waves propagating to the left with
velocity $-2$ remain in the steady state, as shown in
Figure~\ref{patterns}(d). This again agrees with the slope of the
fundamental diagram given by eq. (\ref{r22flux}): in the jammed phase,
$\frac{\partial j}{\partial \rho}=-2$.

The intermediate phase is the most interesting one, because its
dynamics is not dominated by a single type of density waves.
Figures~\ref{patterns}(b) and (c) show two examples of spatio-temporal
patterns in the intermediate phase. One can clearly see that two types
of density waves are present. Blocks containing isolated zeros
propagate to the right, while blocks with isolated ones move to the
left. When these two types collide, they both are slightly delayed,
but after the collision, they continue with the pre-collision
velocity. One could thus say that these density waves are somewhat
similar to solitons -- they preserve their shape and velocity in a
collision.  The balance between these two types of density waves
determines the steady-state current.
As it turns out, this balance not only depends on the density $\rho$,
but also on the amount of correlations present in the initial
configuration. When the initial configuration is described by
Bernoulli probability measure, i.e., all sites are independently
occupied with the same fixed probability, the fundamental diagram is
given by (\ref{r22flux}), and the transition to the intermediate phase
occurs at $\rho_{c1}=\frac{6}{7} -\frac{2\sqrt{2}}{7}$.  Yet the
location of the singularity $\rho_{c1}$ strongly depends on the
assumption of the initial distribution being Bernoulli.

To illustrate this, we prepared ``clustered initial condition'' using
the following algorithm. We start with empty lattice of $L$ sites. In
order to produce initial condition with $\rho L$ occupied sites, we
repeat the following sequence of steps until the number of occupied
sites reaches $\rho L$:
\begin{enumerate}
\item select one site randomly among all empty sites and make it
  occupied by a particle;
\item with probability $q$, move this particle to the site adjacent to
  the nearest occupied site, and with probability $1-q$, leave it in
  the initial position.
\end{enumerate}
In the above, $q \in [0,1]$ is a parameter describing ``clustering''
of occupied sites. When $q=0$, we obtain random non-correlated
configuration. When $q=1$, all occupied sites form a single continuous
block.

\begin{figure}
\begin{center}
\includegraphics[scale=0.9]{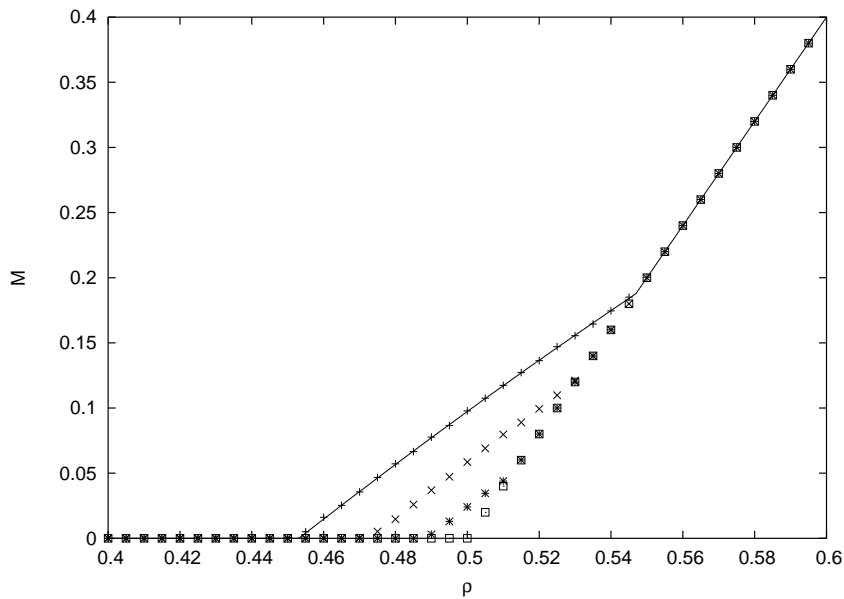}
\end{center}
\caption{Order parameter $M$ as a function of $\rho$
for $q=0$ ($+$), $q=0.2$ ($\times$), $q=0.4$ ($\star$),
and $q=1$ ($\square$). Simulations performed on a lattice of
$10^4$ sites with ${\cal R}_{2,2}$ iterated
$10^4$ time steps. Each point represents average
of $100$ runs.}
\label{corflux}
\end{figure}
Figure~\ref{corflux} shows the order parameter $M$ as a function of
the density $\rho$ for four different values of $q$.  When $q=0$, the
critical point is exactly at $\rho_{c1}$, as expected. However, as $q$
increases, the critical point moves to the right, and when $q=1$, it
reaches $0.5$. Obviously, the same thing happens to the second
singularity, for which one could define analogous order parameter.
When $q=1$, these two singularities merge and we have just one
singularity at $\rho=0.5$. That means that the intermediate phase
disappears as $q \rightarrow 1$.

This behavior is in sharp contrast with the behavior of
number-conserving rules with piecewise-linear fundamental diagrams.
For these rules, the steady-state current depends only on the density,
and does not depend on the amount of spatial correlations present in
the initial condition.

\section{Conclusions}
We investigated properties of the fundamental diagram of rule ${\cal
  R}_{2,2}$. This rule has been used as a representative example of a
class of number-conserving rules for which the fundamental diagram is
known, and for which not all segments of the fundamental diagram are
linear.  We found that the nature of singularities in the fundamental
diagram of ${\cal R}_{2,2}$ is the same as for rules with
piecewise-linear diagrams. The current converges toward its
equilibrium value like $t^{-1/2}$, and the critical exponent $\beta$
is equal to $1$.

These results seem to support a more general conjecture that all
singularities in number-conserving CA rules exhibit universal
behavior. It is interesting to note at this point that the critical
exponent $\beta=1$ is also obtained in equilibrium statistical physics
in the case of second-order phase transitions with non-negative order
parameter and above the upper critical dimensionality. This indicates
that phase transitions in number-conserving CA are all mean-field
type, and the observed universal behavior in in fact mean-field type
behavior.  In order to explore this issue further, it would be
necessary to introduce a field conjugate to the order parameter, and
then numerically compute other critical exponents such as the exponent
$\gamma$ characterizing the divergence of the susceptibility at the
critical point, or $\delta$, characterizing the behavior of the order
parameter at the critical point when the field approaches $0$,
similarly as done in \cite{paper14} for the simplified
Nagel-Screckenberg model.  
This problem is currently under investigation, and results will
be reported elsewhere.

 \vskip 1cm
 \noindent \textbf{Acknowledgements:} The author
acknowledges financial support from NSERC (Natural Sciences and
Engineering Research Council of Canada) in the form of the
Discovery Grant and from SHARCNET in the form of CPU time.

\providecommand{\href}[2]{#2}\begingroup\raggedright\endgroup
\end{document}